\documentclass{vldb}

\usepackage{graphicx}
\usepackage{balance}
\usepackage{moreverb}

\begin{document}

\title{Skew Handling in Aggregate Streaming Queries on GPUs\titlenote{Research conducted while G. Koutsoumpakis and I. Koutsoumpakis were with the Aristotle University of Thessaloniki. The paper was presented at ADMS 2013 - 
Fourth Int. Workshop on Accelerating Data Management Systems Using Modern Processor and Storage Architecture (in conjunction with VLDB 2013).}}

\numberofauthors{3}

\author{Georgios Koutsoumpakis, Iakovos Koutsoumpakis and Anastasios Gounaris}

\author{
\alignauthor
Georgios Koutsoumpakis\\
 \affaddr{Uppsala University}\\
 \affaddr{Sweden}\\
\alignauthor
Iakovos Koutsoumpakis\\
 \affaddr{Uppsala University}\\
 \affaddr{Sweden}\\
\alignauthor
Anastasios Gounaris\\
 \affaddr{Dept. of Informatics}\\
 \affaddr{Aristotle University}\\
 \affaddr{Thessaloniki, Hellas}\\
}

\maketitle
\begin{abstract}
Nowadays, the data to be processed by database systems has grown so large that any conventional, centralized technique is inadequate. At the same time, general purpose computation on GPU (GPGPU) recently has successfully drawn attention from the data management community due to its ability to achieve significant speed-ups at a small cost. Efficient skew handling is a well-known problem in parallel queries, independently of the execution environment. In this work, we investigate solutions to the problem of load imbalances in parallel aggregate queries on GPUs that are caused by skewed data. We present a generic load-balancing framework along with several instantiations, which we experimentally evaluate. To the best of our knowledge, this is the first attempt to present runtime load-balancing techniques for database operations on GPUs.
\end{abstract}

\section{Introduction}

As the amount of data stored and processed in databases increases dramatically, a lot of research effort is being put in the development of advanced techniques that would allow database management systems (DBMSs) to deal with extremely large data volumes more efficiently. Due to the sheer amount of data mentioned above, when conventional, centralized techniques are employed, the cost of processing is raising, thus leading to unacceptable processing times.
In general, a DBMS query response time is affected by two factors, namely the I/O cost, which is the time spent in loading the data from the secondary storage into the main memory, and the computational cost, the time spent by the DBMS while processing the data. Traditionally, research on databases has mostly focused on reducing the I/O cost during query processing, since this is the major bottleneck in many operations. However, the benefits of increasing the system's throughput cannot be ignored as well. So, in an effort to get the most of the database systems, processing throughput should be maximized, and the most broadly established approach to this end is through parallelism. One of the most effective forms of parallelism in queries is partitioned parallelism; partitioned (or intra-operator) parallelism refers to the case where multiple query operators simultaneously process distinct partitions of the same dataset \cite{DG92}.

In this paper, we deal with partitioned queries but we diverge from the conventional DBMS, and we focus on Continuous Query (CQ) systems \cite{MSH+02,flux}. In contrast to traditional DBMSs that answer streams of queries over a non-streaming database, CQ systems treat queries as fixed entities and stream the data over them. This means that the data are not known during the initialization of the query, but new data become available during the execution, so the data items are presented as a possibly infinite sequence of records. Meanwhile, by utilizing a fixed-size sliding window as it typically happens, we ensure that only the most recent data elements are considered when answering queries. In this way, the query results depend only on the latest and most up to date data. CQs are particularly applicable to streaming scenarios, where data is produced at such a fast rate that it is not practical first to store  the data and then to process it; rather, data processing must be performed on the fly. More specifically, we investigate parallel aggregate queries over data streams, i.e., queries that split the dataset into groups, and for each group continuously update the value of an aggregate function leveraging the group-by database operator \cite{GM08}. E.g., in a stream of stock market data, we can pose a query to continuously output the average price in the last one thousand transactions for each stock. Such queries are essential in challenging, real-time data-mining applications \cite{HG09,CDN11}.

An efficient way of increasing the computational capacity can be achieved by taking advantage of the parallelization capabilities and high computational power offered by modern graphics processing units (GPUs). This is commonly referred to as general purpose computing on GPUs (GPGPU).
Since GPUs are especially designed for stream processing and provide free programmable processing cores, they are well suited to be used for several database operations, for example group-by query execution. Many-core technologies like NVidia's Compute Unified Device Architecture (CUDA) and the associated Fermi hardware architecture \cite{nvidia,nvidia-fermi} that have been built on top of GPUs simplify the development of highly parallel algorithms running on a single GPU. In general, a GPU can only execute algorithms for processing the data, called kernels, while the corresponding control logic is executed on the CPU. Previous research in the field has led to the implementation of database management systems that allocate data intensive tasks to the GPU. Examples of such systems are Sphyraena \cite{sfyraena} and GPUQP \cite{HLY+09}, which are implementations of DBMSs for GPUs.

At the initialization of a CQ query, the query engine selects a default configuration, which provides the settings about how the query will execute among the graphics card's threads. Queries on a data stream will, by definition, run long enough to experience changes in data properties as well as system workload during their run. This implies that workload imbalances among the threads may occur, causing bottleneck to the system if the workload assignment is skewed, i.e., some processing units receive more work than the others. In parallel aggregate queries, the main cause of skewed execution is due to skewed data value distributions, because the assignment of data groups to processing units (which are GPU threads in our case) depends on data values, the distribution of which may be volatile. A continuous query engine should adapt gracefully to these changes or correct any bad initial workload assignments at runtime, in order to ensure efficient processing over time. To this end, runtime load balancing is employed, as data processing is dynamically reassigned among the card's threads.

In this paper, we investigate solutions to the problem of skewed execution in aggregate queries on GPUs, where the CPU and the GPU closely cooperate to achieve high performance. More specifically, we present a load balancing framework, where a load-balancing coordinator runs on the CPU to decide the allocation of groups to threads, whereas the execution of the query takes place on the GPU. Then, we introduce a family of load-balancing policies that instantiate the framework and we experimentally evaluate them. To the best of our knowledge, this is the first attempt to present runtime load-balancing techniques for database operations on GPUs.

The remainder of this article is structured as follows: the next section discusses the related work. Section \ref{sec:arch} deals with the runtime load-balancing architecture. The detailed approaches to load-balancing are presented in Sec. \ref{sec:techniques}. In Section \ref{sec:eval},  we evaluate the efficiency in skew handling for each of the proposed approaches. We conclude in Section \ref{sec:conclusions}.

\section{Related Work}
\label{sec:rw}

An increasing number of researchers and practitioners use GPUs instead of CPU clusters for data- and computation-intensive problems. The proposals that are most closely related to our work include those that refer to the development of query processing techniques on GPUs. Although there are several early efforts towards this research direction (e.g., \cite{Bandi:2004}), only recently fully-fledged query processing systems have appeared. The two most prominent examples are SphyraEna \cite{sfyraena} and GPUQP \cite{HLY+09,HYF+08}, which feature a fully functional DBMS with the capability to execute queries on the GPU, in order to benefit from its computing power. GPUQP provides a query engine where the queries (containing operators such as join, group-bys, and so on) run either entirely on the GPU or, in some cases, on both the CPU and the GPU. The database is not stored in the card's memory, but on the disk, as in conventional databases. When a query is to be executed, the system employs techniques to estimate the total cost, which includes data transfer to the GPU's memory and computational cost, and then to decide which parts of the query plan should be allocated to the GPU. Our work is different in the sense that we focus on a specific part of queries, namely aggregate queries, and we deal with the runtime load-balancing problem that is not considered by systems such as Sphyraena and GPUQP.

In addition, the MapReduce programming framework has emerged as an alternative environment for processing large amounts of data \cite{DG08}. MapReduce inherently supports aggregate queries. The map phase splits data into groups, and the reduce phase is responsible for computing the aggregate function for each resulting group. An implementation of the MapReduce paradigm on GPUs has been proposed in \cite{FangHLG11}. Nevertheless, in MapReduce, no dynamic load balancing that modifies the allocation of reducers on the fly is supported.

A load balancing proposal for GPUs has appeared in \cite{Cederman:2008}. The setting assumed by this work is quite different from the one in a streaming query though, since it relies on queues holding tasks. Each processor maintains a queue that contains the tasks to perform. When a task is carried out, it is popped from the queue and the processor deals with the next one. As a method of load balancing, when a processor is idle, it tries to steal tasks from the next processor's queue tail until all the tasks are completed. By contrast, in our work we do not employ queues for tasks, but each data group is assigned to a specific thread and decisions may be revised in each iteration, as explained in the following sections. Note also that our work is orthogonal to proposals that aim to fine tune GPUs in order to maximize performance (e.g., \cite{Sor11}).

Concerning stream processing and load balancing in CQ systems, a lot of available research material proposes methods for rebalancing data distribution to the processors \cite{WYT+93,RM95,Rah96,RM93,GounarisTAAS12,flux,NormanVldb08}. Generally, the producer-consumer model is applied, where producers simply perform the data distribution and delegate the data processing to the consumers, who are responsible for the execution of the processing logic. Usually, the consumers do not have the same throughput, and the slowest of them acts as a bottleneck to the CQ system. Moreover, the data properties often change with time, causing uneven allocation and the need to redefine the query execution. In the Flux model \cite{flux}, two methods of load balancing are proposed: the short term balancing method and the long term one. In the former case, a buffer for each consumer ensures that no producers will have to stay suspended until the consumer they want to deal with finishes his former work. As this method is inadequate for dealing with long term workload imbalances, Flux introduces a mechanism for long term rebalancing using state transfer. As data is divided in small partitions and distributed to the processors, when imbalance is detected, a partition is dynamically relocated to another processor on the fly. Flux techniques have been extended and improved upon in \cite{GounarisTAAS12,NormanVldb08}. Our work can be deemed as a proposal for techniques for long term imbalances in a GPU setting.

In general, techniques that modify the query execution at runtime are commonly referred to as adaptive query processing (AQP) ones \cite{Deshpande-07}. AQP for CQs may have several additional flavors. For example, the proposals in \cite{MSH+02,ZRH04} try to effectively solve the load balancing problem
by dynamically changing the query execution plan. More specifically, if multiple joins are being executed, the techniques developed  change their  order at runtime with a view to reducing the total processing cost. Other adaptive proposals that refer to CQs but do not deal with re-partitioning issues are discussed in \cite{KKP11,PMCL11}.

\begin{figure*}[tb!]
\centering
  \includegraphics[width=0.92\textwidth]{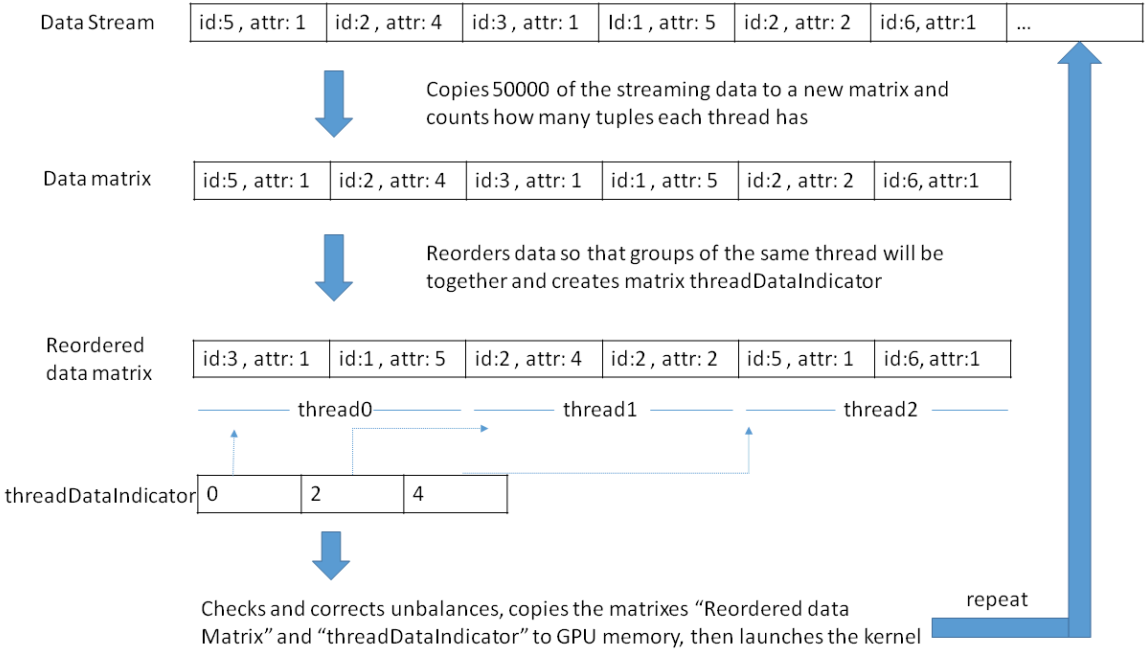}\\
  \caption{High level data flow and CPU operations.}\label{fig:cpu}
\end{figure*}

\section{Our Load-balancing Framework}
\label{sec:arch}


\begin{figure*}[tb!]
\centering
  \includegraphics[width=0.85\textwidth]{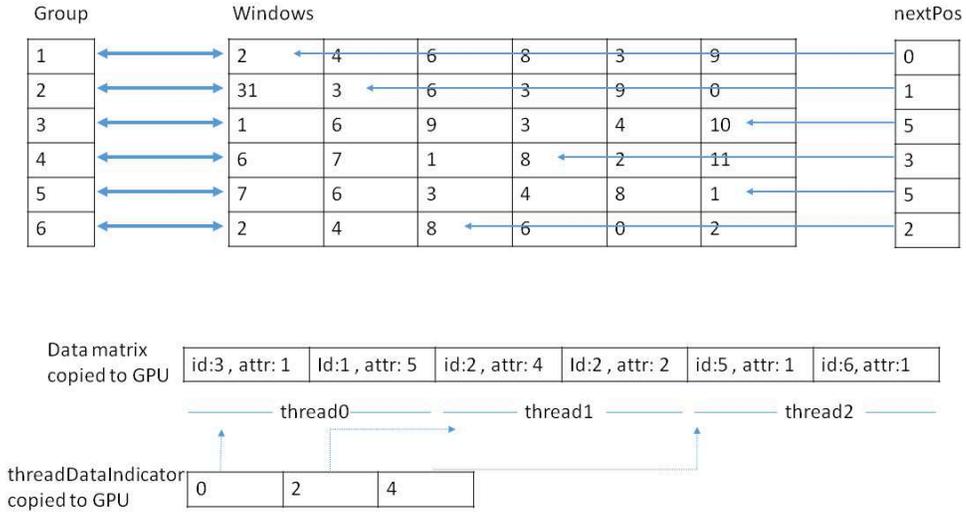}\\
  \caption{Auxiliary data structures on the GPU.}\label{fig:gpu}
\end{figure*}

In this section we describe the architecture and the high-level approach to load balancing  for aggregate queries on GPUs; the exact balancing algorithms are presented in Section \ref{sec:techniques}. As the GPU is only suitable for executing kernels and the logic of the program is administered by the CPU, extensive collaboration between the two units is required. The CPU is responsible (i) for preparing the data so that coalesced memory access on the CUDA is enabled and (ii) for detecting and correcting imbalances. The GPU takes over the actual data processing. To this end, some auxiliary structures on both processing units are created in an effort to maximize the throughput of the system.

Throughout this paper, we consider a scenario where we execute a group-by query over data that consist of two fields, namely a group identifier {\it id} and an integer value {\it attr}. For each new tuple, the aggregate function needs to process the last attribute values received for each group identifier according to a fixed sized window; i.e., the aggregate value depends on the recent history of the stream. Although this is a specific scenario, it possesses all the characteristics so that it can easily be extended and generalized to further demanding streaming aggregate queries.

\subsection{Operations on the CPU}
As the data arrives at the CPU as a stream, the processing takes place in iterations. An example data and control flow is shown in Figure \ref{fig:cpu}. In each iteration, a fixed size batch of tuples is processed, e.g., in the example in Figure \ref{fig:cpu} the batch size is 50K. The CPU maintains two auxiliary structures that (i) map groups to specific GPU threads that will process them and, (ii) map, in the reverse way, GPU threads to groups (these structures are not shown in the figure).  Using the former, it organizes the data in a matrix, which will be later copied to the GPU's global memory, in a way that all the tuples of the groups allocated to the same thread will be in adjacent memory slots to allow for coalesced memory access; i.e., the matrix is not fully sorted. The {\it Reordered Data matrix} in Figure \ref{fig:cpu} illustrates an example. This matrix is produced in linear time, and more specifically, in two passes. In the first pass, for each thread, we count the occurrences of the data items that belong to groups assigned to that thread. This provides adequate information about the exact places in the matrix, where each data item should be placed, given that we know the mapping of groups to threads. The actual placement takes place in the second pass.

The CPU also utilizes an array  that indicates where each GPU's thread should seek for its assigned data during the kernel launch; this is called {\it threadDataIndicator}.

Before copying the data to the GPU and calling the kernel, the skew handling techniques intervene, in order to investigate whether any imbalance occurs. The rebalancing technique used might then propose a new mapping of a group to a different thread. So, the group-to-thread mapping structures are continuously updated. However, since it is expensive to revise the ordered data matrix and its associated thread indicator array, the effects of the rebalancing take effect from the next iteration; in other words there is a delay of one iteration in our skew handling approach.

At the end of each iteration, the {\it Reordered Data matrix} and the {\it threadDataIndicator} array are copied to the GPU's global memory. At this point, the kernel is launched asynchronously as a single CUDA stream. With the help of these two data structures that are copied to the GPU, we ensure that every thread will be capable of instantly locating the data it is responsible for. Note that in GPGPU, each thread can be uniquely identified in a straightforward manner.

While waiting for the GPU to finish with the processing of the batch of tuples sent, the CPU prepares the next batch of streaming data. As explained above, for the next batch, the allocation of groups to threads will be based on the group-to-thread mappings that have resulted from the previous batch.

It is important to mention that, as verified also by our experiments, the preparation time on CPU overlaps with the more expensive GPU operations. When the grid size is small, the overlap is full and thus the load-balancing overhead is hidden. However, when the grid size increases, the overlap is partial and the overhead has an impact on the total running time.

\subsection{Operations on the GPU}

Upon receipt of a new batch of ordered tuples, each thread starts processing tuples. To do so, it accesses only the relevant data matrix cells (with the help of the {\it threadDataIndicator} array). For each tuple, the GPU thread adds it to a persistent data structure in the global memory that holds all the windows for all groups. If the window is full, the oldest tuple of the window has to be discarded. Overall, the auxiliary data structures on the GPU are: (i) a matrix structure that keeps the windows for all the groups, (ii) an array that maps each group to its window in the previous matrix,  and (iii) an array structure that contains pointers to the oldest value of the window for each group ({\it nextPos}) (see Figure \ref{fig:gpu}). When the oldest tuple is replaced, the pointer moves to the next cell in the same window.

To compute the aggregate function, each thread, for each allocated group,  processes all the tuples within the window. When all the tuples have been processed, the kernel call terminates. Note that no new kernel can be called before termination. This implies that no data on the GPU are replaced with new data before their processing, thus ensuring correctness. As shown by the experiments in the next section, such an approach is not only capable of skew handling, but can also increase the system throughput significantly. However, there may still exist cases, where the data arrival rate exceeds the GPU capacity; in that case, our proposal needs to be complemented with load shedding and/or approximate techniques, which we leave for future work.

\section{Load-balancing Techniques}
\label{sec:techniques}

\begin{figure*}[tb!]
\begin{center}
{\small
\begin{tabbing}
aaa \= aa \= aaa \= aaa \= aaa \= aaa \kill
\rule{\columnwidth}{0.5pt}\\
{{\bf Algorithm} getFirst~($\vec{tpt}$, $threadThreshold$)}\\
\> $\vec{tpt}$: a vector containing the number of tuples per thread,\\
\> $threadThreshold$: threshold to indicate imbalance\\
\rule{\columnwidth}{0.2pt}\\
1. \> Find the thread with the highest number of tuples assigned $tmax$; \\
2. \> Find the thread with the lowest number of tuples assigned $tmin$; \\
3. \> {\bf While} $tpt[tmax]-tpt[tmim]>threadThreshold$; \\
4. \>\> Assign the first group of thread $tmax$ to the thread $tmin$; \\
5. \>\> Find the new $tmax$,$tmin$;\\
6. \> {\bf endWhile};\\
\rule{\columnwidth}{0.5pt}\\
\end{tabbing}
} \caption{Outline of the {\it getFirst} algorithm.} \label{alg1}
\end{center}
\end{figure*}

\begin{figure*}[tb!]
\begin{center}
{\small
\begin{tabbing}
aaa \= aa \= aaa \= aaa \= aaa \= aaa \kill
\rule{\columnwidth}{0.5pt}\\
{{\bf Algorithm} checkAll~($\vec{tpt}$, $threadThreshold$)}\\
\> $\vec{tpt}$: a vector containing the number of tuples per thread,\\
\> $threadThreshold$: threshold to indicate imbalance\\
\rule{\columnwidth}{0.2pt}\\
1. \> Find the thread with the highest number of tuples assigned $tmax$; \\
2. \> Find the thread with the lowest number of tuples assigned $tmin$; \\
3. \> {\bf While} $tpt[tmax]-tpt[tmim]>threadThreshold$; \\
4. \>\> Read the tuples of thread $tmax$ and find the group with the most occurrences; \\
5. \>\> Assign that group to the thread $tmin$; \\
6. \>\> Find the new $tmax$,$tmin$;\\
7. \> {\bf endWhile};\\
\rule{\columnwidth}{0.5pt}\\
\end{tabbing}
} \caption{Outline of the {\it checkAll} algorithm.} \label{alg2}
\end{center}
\end{figure*}

\begin{figure*}[tb!]
\begin{center}
{\small
\begin{tabbing}
aaa \= aa \= aaa \= aaa \= aaa \= aaa \kill
\rule{\columnwidth}{0.5pt}\\
{{\bf Algorithm} probCheck~($\vec{tpt}$, $threadThreshold$, $pot$)}\\
\> $\vec{tpt}$: a vector containing the number of tuples per thread,\\
\> $threadThreshold$: threshold to indicate imbalance,\\
\> $pot$: thepercentage of thread's tuples that the to-be-reassigned group has to cover\\
\rule{\columnwidth}{0.2pt}\\
1. \> Find the thread with the highest number of tuples assigned $tmax$; \\
2. \> Find the thread with the lowest number of tuples assigned $tmin$; \\
3. \> {\bf While} $tpt[tmax]-tpt[tmim]>threadThreshold$; \\
4. \>\> $ngroups$ = number of groups assigned to $tmax$; \\
5. \>\> $limit$ = $pot \ast tpt[max]/ngroups$; \\
6. \>\> Read the tuples of thread $tmax$ and stop when a group appears $limit$ times; \\
7. \>\> Assign that group to the thread $tmin$; \\
8. \>\> Find the new $tmax$,$tmin$;\\
9. \> {\bf endWhile};\\
\rule{\columnwidth}{0.5pt}\\
\end{tabbing}
} \caption{Outline of the {\it probCheck} algorithm.} \label{alg3}
\end{center}
\end{figure*}

\begin{figure*}[tb!]
\begin{center}
{\small
\begin{tabbing}
aaa \= aa \= aaa \= aaa \= aaa \= aaa \kill
\rule{\columnwidth}{0.5pt}\\
{{\bf Algorithm} bestBalance~($\vec{tpt}$, $threadThreshold$)}\\
\> $\vec{tpt}$: a vector containing the number of tuples per thread,\\
\> $threadThreshold$: threshold to indicate imbalance\\
\rule{\columnwidth}{0.2pt}\\
1. \> Find the thread with the highest number of tuples assigned $tmax$; \\
2. \> Find the thread with the lowest number of tuples assigned $tmin$; \\
3. \> {\bf While} $tpt[tmax]-tpt[tmim]>threadThreshold$; \\
4. \>\> Read the tuples of thread $tmax$ and find the group that, if swapped, minimizes $tpt[tmax]-tpt[tmin]$; \\
5. \>\> Assign that group to the thread $tmin$; \\
6. \>\> Find the new $tmax$,$tmin$;\\
7. \> {\bf endWhile};\\
\rule{\columnwidth}{0.5pt}\\
\end{tabbing}
} \caption{Outline of the {\it bestBalance} algorithm.} \label{alg4}
\end{center}
\end{figure*}

\begin{figure*}[tb!]
\begin{center}
{\small
\begin{tabbing}
aaa \= aa \= aaa \= aaa \= aaa \= aaa \kill
\rule{\columnwidth}{0.5pt}\\
{{\bf Algorithm} shift~($\vec{tpt}$, $threadThreshold$)}\\
\> $\vec{tpt}$: a vector containing the number of tuples per thread,\\
\> $threadThreshold$: threshold to indicate imbalance\\
\rule{\columnwidth}{0.2pt}\\
1. \> Find the thread with the highest number of tuples assigned $tmax$; \\
2. \> Find the thread with the lowest number of tuples assigned $tmin$; \\
3. \> {\bf While} $tpt[tmax]-tpt[tmim]>threadThreshold$; \\
4. \>\> {\bf if} $tmax>tmin$ \\
5. \>\>\> {\bf foreach} thread $i \in (tmin,tmax]$\\
6. \>\>\>\> Move the first group from thread $i$ to the thread $i-1$; \\
7. \>\> {\bf else} \\
8. \>\>\> {\bf foreach} thread $i \in [tmax,tmin)$\\
9. \>\>\>\> Move the last group from thread $i$ to the thread $i+1$; \\
10. \>\> Find the new $tmax$,$tmin$;\\
11. \>\> {\bf endif} \\
12. \> {\bf endWhile};\\
\rule{\columnwidth}{0.5pt}\\
\end{tabbing}
} \caption{Outline of the {\it shift} algorithm.} \label{alg5}
\end{center}
\end{figure*}

\begin{figure*}[tb!]
\begin{center}
{\small
\begin{tabbing}
aaa \= aa \= aaa \= aaa \= aaa \= aaa \kill
\rule{\columnwidth}{0.5pt}\\
{{\bf Algorithm} shiftLocal~($\vec{tpt}$, $threadThreshold$)}\\
\> $\vec{tpt}$: a vector containing the number of tuples per thread,\\
\> $threadThreshold$: threshold to indicate imbalance\\
\rule{\columnwidth}{0.2pt}\\
1. \> {\bf foreach} thread $i$\\
2. \>\> {\bf if}  $tpt[i]-tpt[i+1]>threadThreshold$; \\
3. \>\>\> Move the last group from thread $i$ to the thread $i+1$; \\
4. \>\> {\bf else if} $tpt[i+1]-tpt[i]>threadThreshold$; \\
5. \>\>\> Move the first group from thread $i+1$ to the thread $i$; \\
\rule{\columnwidth}{0.5pt}\\
\end{tabbing}
} \caption{Outline of the {\it shiftLocal} algorithm.} \label{alg6}
\end{center}
\end{figure*}

As we have previously mentioned, the number of tuples each thread has to process is calculated on the CPU before the kernel call. At that stage, it is easy to identify imbalances. In all the re-balancing methods except the last one, we follow the same pattern: we keep two heaps, a min heap and a max heap, which contain information about the most and least loaded threads, respectively (in $O(1)$ time). Then, we match the most and the least loaded thread into a pair.  To indicate imbalance, we introduce a threshold of tuple count difference. If the difference is above the threshold, a group is moved from the most loaded thread to the least loaded one, depending on the technique selected, and the mapping structures are updated. Then, we execute the same process for the new most and least loaded threads. Since the number of this type of iterations may be high, the choice of heap data structures is beneficial.  The methods for choosing the group that will be moved in case of imbalance are explained below.

{\it getFirst} is the least sophisticated technique and requires only small computation effort. The group to be moved is the first one in the thread-to-group mapping structure. The technique is given in Figure \ref{alg1}. In {\it checkAll} (see Figure \ref{alg2}), the group to be moved is the one with the most appearances in the thread. All the tuples of the thread in the current batch have to be scanned, in order to count the group appearances. Then, the most frequent group is selected and remapped with a view to correcting the imbalance faster than the previous technique.

{\it probCheck} is as an approximate version of {\it checkAll}, which tries to locate the most common group without having to scan all the tuples. To manage this, it first calculates the average number of tuples in the groups of the most loaded thread. Then, it selects the first group detected with frequency equal to $pot$  times that average value ($0<pot \leq 1$). The higher {\it pot} is, the higher the chance is the most common group will be selected at the expense of increased data scanning cost; however, the scanning cost is always less than the cost of scanning all the thread's tuples. {\it probCheck} is shown in Figure \ref{alg3}.

The algorithms thus far make simple choices as to which groups should be allocated to other threads. Neither random choices nor selecting the most frequent group can guarantee that the imbalance will be eliminated across all threads. {\it bestBalance} tries to address this limitation and detects the group that, if remapped, will achieve the best balance between the two corresponding threads. To achieve this, it scans all the tuples assigned to the most loaded thread, counting the appearances of each group. Then, it chooses the group that minimizes the difference in the workload, as shown in Figure \ref{alg4}.

The last two methods aim to further benefit from coalesced memory access. More specifically, the {\it shift} method inserts a locality criterion in the skew handling according to which the groups are not moved directly from one thread to another, but only to the neighboring one. There are two cases: if the loaded thread has a smaller id number than the emptier one, then each thread in the range $[loaded, unloaded)$ has its last group assigned to the next thread. Consequently, the loaded thread will have to process one group less, while the least loaded one is now assigned with some extra load. The same happens if the loaded thread has a bigger id, only the other way around (see Figure \ref{alg5}).

{\it shiftLocal} does not rely on the detection of the most and least loaded threads, thereafter it does not require the two heaps. {\it shiftLocal} only fixes imbalances among neighboring  threads. It compares each thread's load to the next one's, using an appropriate threshold factor, and properly moves the last or first group to the less loaded thread (see Figure \ref{alg6}).

\section{Evaluation}
\label{sec:eval}

In this section, we evaluate the efficiency and the effectiveness of our approach to skew handling. We focus on both performance improvements and the associated overheads. For completeness, we examine scenarios with no, low and high imbalance.

\subsection{Experimental Setting}

For the purpose of our experiments, two different system configurations supporting the Fermi architecture are used: the first system (referred to as {\it PC1}) has an Intel Core2 Duo E6750 CPU at 2.66GHz and an NVidia 460GTX (GF104) graphics processor at 810 Mhz on a PCIe v2.0 x16 slot (5GB/s transfer rate). {\it PC2} has an Intel P4 550 CPU, running at 3.4 Ghz. Also, it has an NVidia 550GTX Ti (GF116) at 910 MHz, which is installed on a PCIe v1.1 x16 (2.5GB/s transfer rate) slot. In both cases, our techniques have been developed using the NVidia Parallel NSight 2.1 platform. The two configurations are appropriately selected to favor the investigation of the relative performance of both (i) a system with slower CPU but more powerful GPU and (ii) a system with a faster CPU but slower GPU.

Also, we experiment with three datasets, namely {\it DS1, DS2} and {\it DS3}. {\it DS1} consists of unskewed data. It comprises 100M tuples, assigned to 40000 groups in a round robin way, so that the allocation of tuples to groups follows a uniform distribution. This dataset does not require any runtime balancing and is used for comparison purposes. {\it DS2} follows a zipf distribution. It consists of 100M tuples as well.
In {\it DS2} the group ids are assigned in such a way that a group with id equal to $y$  is more frequent than the groups with id $z$, if $z>y$. Finally, {\it DS3} is a randomly permuted version of the {\it DS2}, so that the group ids are not in decreasing order of frequency.

In each iteration, the batch of tuples consists of 50K tuples. As such, in our experiments, the processing finishes after 2,000 iterations for all datasets, but the results can be transferred to infinite streams as well. The size of the sliding window that needs to be maintained for each group is 100 tuples, and, initially each thread receives an equal number of groups with consecutive group ids. After every new tuple is copied to the appropriate window of a group, the complete window is scanned and its sum is calculated from scratch thus simulating a demanding data analysis task. When executing the kernel, we set the block size to 256 threads, but we vary the grid size. The {\it threadThreshold} value is set to 1000. For the {\it probCheck} method, the {\it pot} parameter is set to 0.5, which was experimentally found as the optimal value. All experiments were conducted three times and the average value is presented. The standard deviation is depicted when it is not negligible; in general it is very small.

\subsection{Experiments}

\subsubsection{Performance degradation due to imbalance}
\begin{figure}[tb!]
\begin{center}
\includegraphics[width=0.46\textwidth]{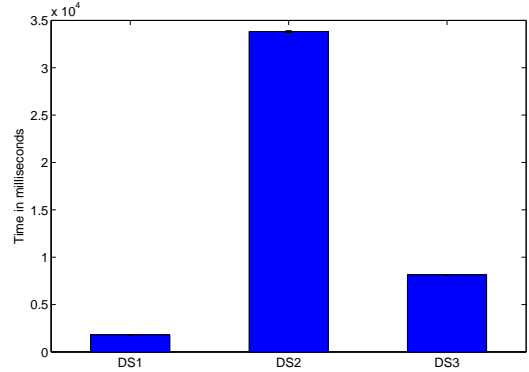}
\caption{Comparison of execution without load-balancing in PC1 for grid size 4.}
\label{fig1}
\end{center}
\end{figure}

We start our experiments by showing the detrimental effects of not performing load balancing when data is skewed. In Figure \ref{fig1}, the left column that corresponds to DS1 demonstrates the performance of our aggregation operator in the optimal case: dynamic balancing is neither needed nor performed (i.e., there is no overhead). However, for DS2, if no balancing technique is activated, the execution time increases by an order of magnitude, even if the total size to be processed remains the same.
This is because only a few threads are burdened with the majority of the processing, while the others remain idle. In the case of DS3, although the allocation of tuples in groups is still skewed, the most common groups are randomly distributed among the threads, rather than being allocated to the first ones, so the overall increase in the execution time is smaller compared to DS2. In general, DS2 is regarded as a scenario, where the imbalance is high, and DS3 corresponds to a scenario with lower imbalance.

\subsubsection{Performance improvements}

\begin{figure*}[tb!]
\begin{center}
\begin{tabular}{cc}
\includegraphics[width=0.49\textwidth]{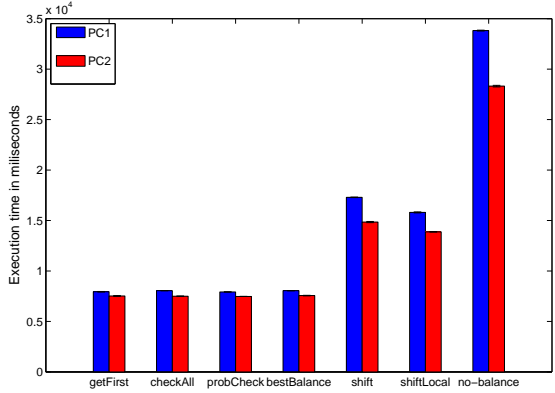} & \includegraphics[width=0.49\textwidth]{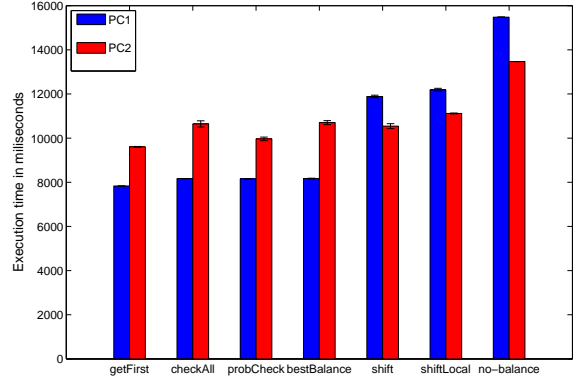} \\
\end{tabular}
\caption{Comparison of execution time of load-balancing methods for DS2 (high imbalance) with grid size 4 (left) and 64 (right).}
\label{fig:DS2}
\end{center}
\end{figure*}

In Figure \ref{fig:DS2}, we present the evaluation results of the six methods of skew handling that were discussed in Section \ref{sec:techniques} for DS2 (high imbalance), and we compare them against the case where we do not perform load balancing. As expected, runtime skew handling results in considerable improvement in the execution of parallel aggregate queries execution, since the workload is more evenly distributed among the graphic card's processor units.
However, as indicated in the figures, not all the methods can achieve the same improvement. More specifically, {\it shift} and {\it shiftLocal} are inferior to the other four methods for this dataset. This is due to the fact that, while these methods handle the rebalancing by moving groups to neighboring threads, in the DS2 dataset, the major amount of tuples is distributed among the first threads. Therefore, in order to maintain an acceptable level of balance, {\it shift} and {\it shiftLocal} require many iterations in order to fix the imbalance.

Comparing the other methods, we can draw the following observations: {\it probCheck} is slightly more effective than {\it checkAll}, as it chooses faster, albeit in a probabilistic manner, the optimal group to move, without having to consider all the tuples of the stream. On the other hand, the {\it bestBalance} method guarantees the optimal solution for the skew handling problem in terms of equalizing the workload among GPU threads. This implies that it requires the least group repartitioning effort in future batches. Nevertheless, as the balancing decisions are enforced after one round, where the conditions might have changed, and the demand for CPU processing time in order to compute the best solution increases, in most of the cases, the execution time of {\it bestBalance} is slightly higher than those of {\it getFirst, checkAll} and {\it probCheck}. Interestingly, as {\it getFirst} arbitrarily moves the first group of the most loaded thread, it cannot deal with severe imbalances successfully in a few rebalancing iterations, but because of its low overhead, it leads to competitive query execution time.

\begin{figure*}[tb!]
\begin{center}
\begin{tabular}{cc}
\includegraphics[width=0.49\textwidth]{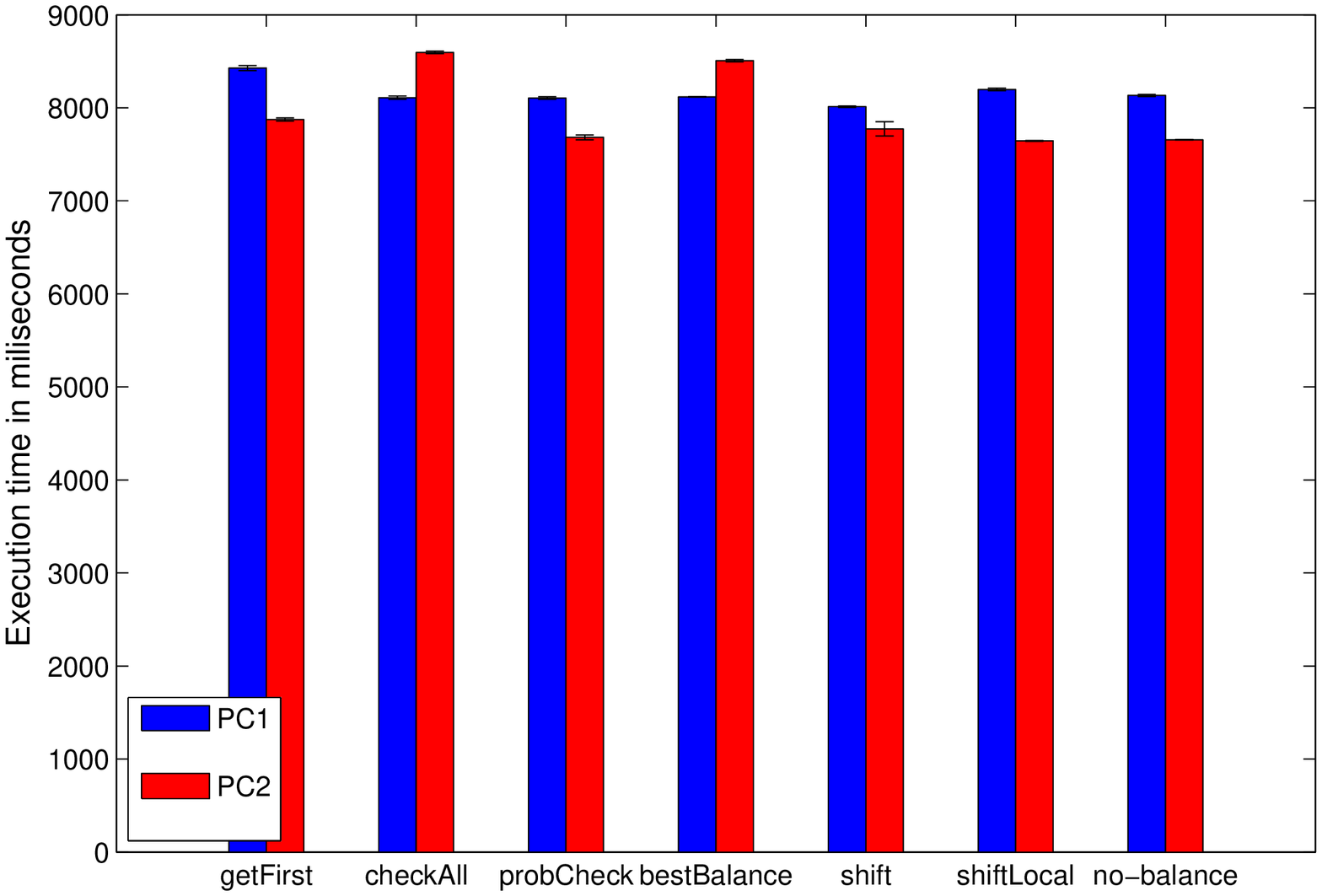} & \includegraphics[width=0.49\textwidth]{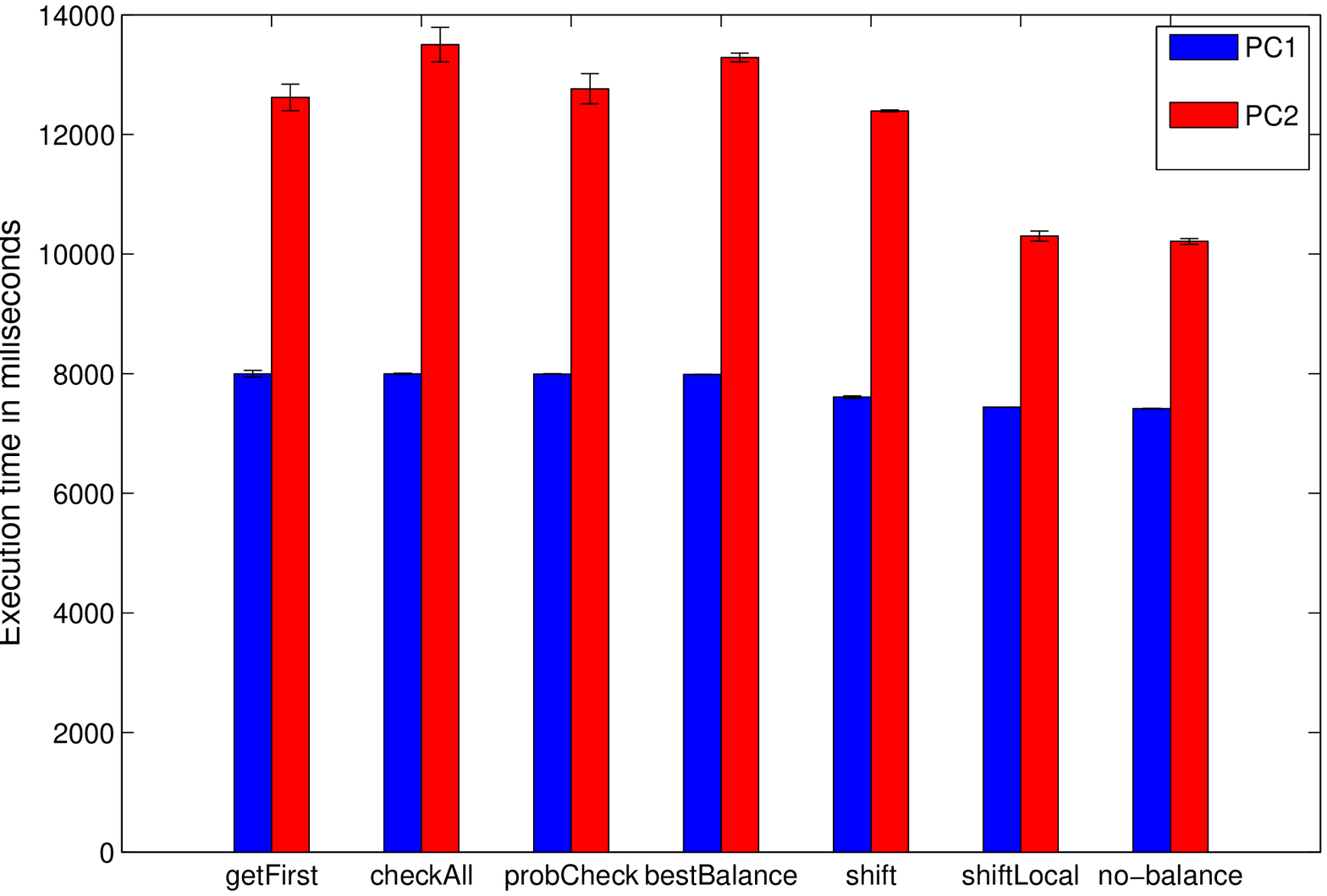} \\
\end{tabular}
\caption{Comparison of execution time of load-balancing methods for DS3 (low imbalance) with grid size 4 (left) and 64 (right).}
\label{fig:DS3}
\end{center}
\end{figure*}

\begin{table}[tb!]
\centering
\scriptsize
\begin{tabular}{|c||p{0.07\textwidth}|p{0.07\textwidth}||p{0.07\textwidth}|p{0.07\textwidth}|}
  \hline
  & \multicolumn{2}{c||}{High Imbalance (DS2)}   & \multicolumn{2}{c|}{Low Imbalance (DS3)}   \\  \hline
Technique & PC1 & PC2 & PC1 & PC2 \\ \hline \hline
getFirst & 4.26 &	3.76 &	0.97 & 0.97 \\ \hline
checkAll & 4.20 &	3.78 &	1 &	0.89 \\ \hline
probCheck & {\bf 4.27} &	{\bf 3.79} &	1 &	{\bf 1} \\ \hline
bestBalance & 4.20 & 3.75 &	1 &	0.9 \\ \hline
shift & 1.96 &	1.91 &	{\bf 1.01} &	0.99 \\ \hline
shiftLocal & 2.14 &	2.04 &	0.99 &	{\bf 1} \\ \hline
no balance & 1 &	1 &	1 &	1 \\ \hline
\end{tabular}
\caption{Normalized throughput (tuples processed in time units) for grid size 4.}
\label{Table:exp1-time}
\end{table}

\begin{table}[tb!]
\centering
\scriptsize
\begin{tabular}{|c||p{0.07\textwidth}|p{0.07\textwidth}||p{0.07\textwidth}|p{0.07\textwidth}|}
  \hline
  & \multicolumn{2}{c||}{High Imbalance (DS2)}   & \multicolumn{2}{c|}{Low Imbalance (DS3)}   \\  \hline
Technique   & PC1        & PC2         & PC1 & PC2 \\ \hline \hline
getFirst    & {\bf 1.98} &	{\bf 1.4}  &	0.93            & 0.81 \\ \hline
checkAll    & 1.9        &	1.27       &	0.93            &	0.76 \\ \hline
probCheck   & 1.9        &	1.35       &	0.93            &	0.8 \\ \hline
bestBalance & 1.9       & 1.26         &	0.93             &	0.77 \\ \hline
shift       & 1.3       &	1.28       &	0.97              &	0.82 \\ \hline
shiftLocal  & 1.27      &	1.21       &	{\bf 1}         &	{\bf 0.99} \\ \hline
no balance  & 1         &	1          &	1            &	1 \\ \hline
\end{tabular}
\caption{Normalized throughput (tuples processed in time units) for grid size 64.}
\label{Table:exp1-time64}
\end{table}

\begin{figure*}[tb!]
\begin{center}
\begin{tabular}{cc}
\includegraphics[width=0.48\textwidth]{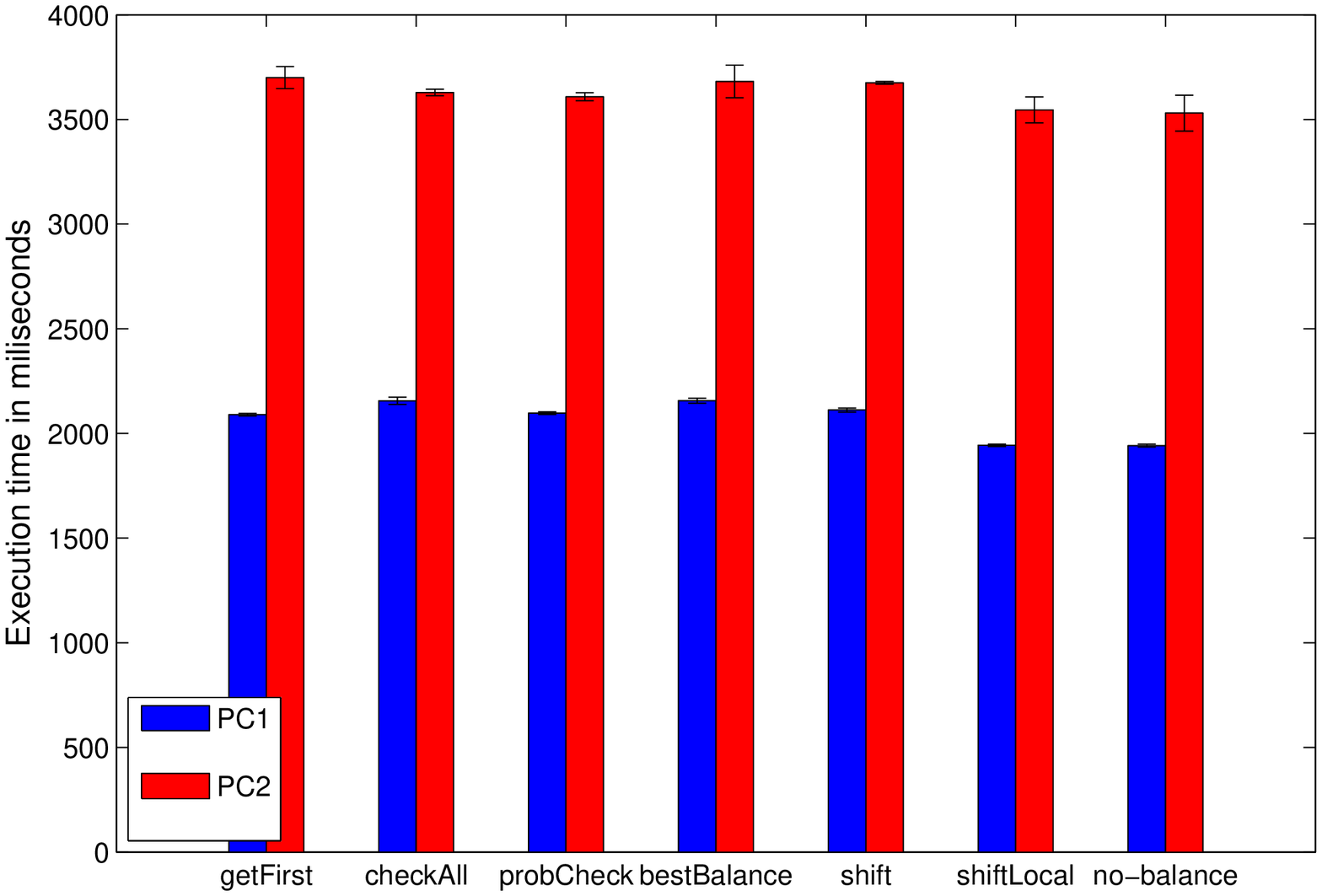} & \includegraphics[width=0.48\textwidth]{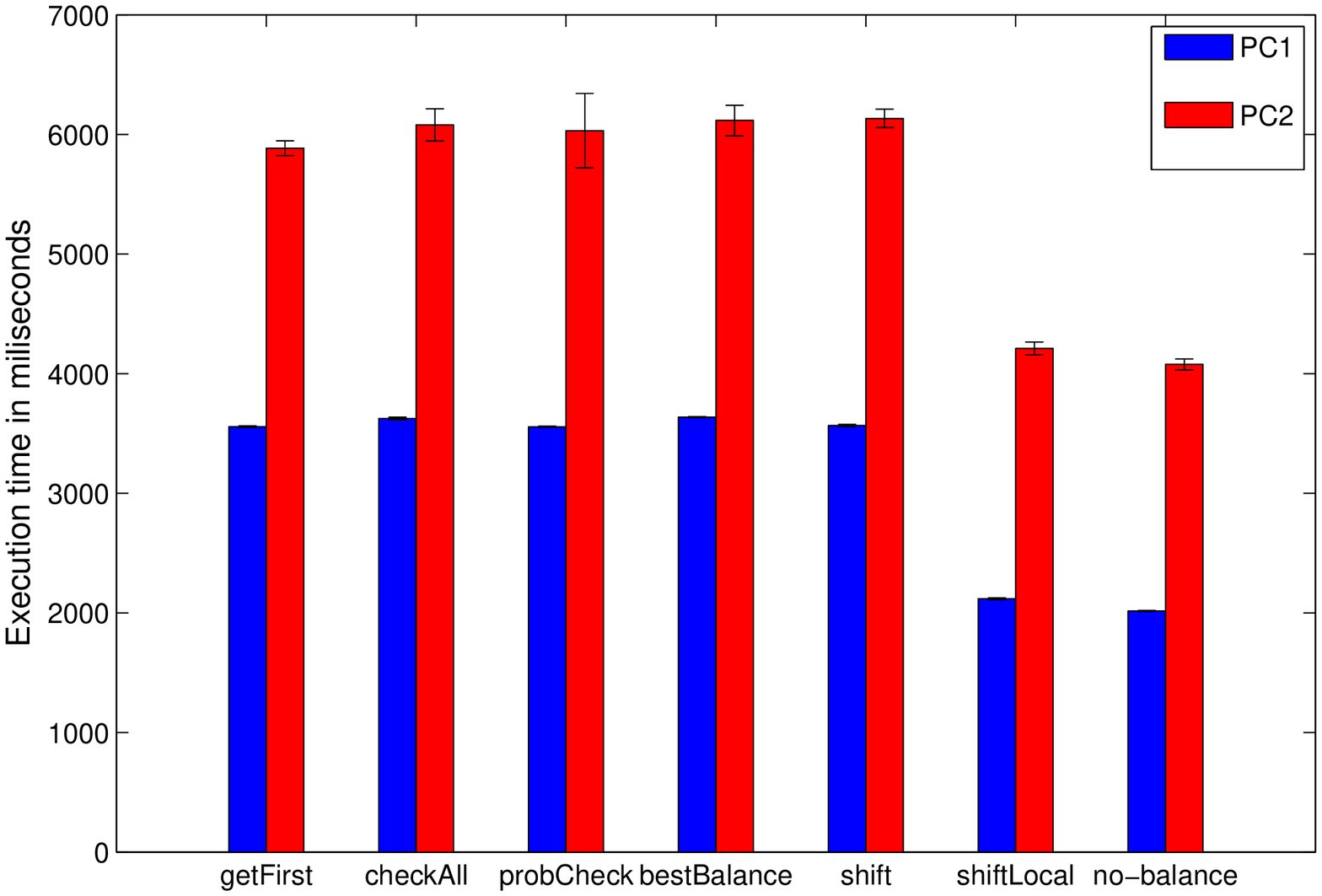} \\
\end{tabular}
\caption{Comparison of execution time of load-balancing methods for DS1 (no imbalance) with grid size 4 (left) and 64 (right).}
\label{fig:DS1}
\end{center}
\end{figure*}

As mentioned earlier, DS2 corresponds to a scenario with high imbalance. In DS3, the imbalance is lower, and also, the loaded threads may have arbitrary ids. We repeat the same experiment for DS3, and the results appear in Figure \ref{fig:DS3}. There are two main observations: firstly, no load balancing technique is actually effective, and secondly, {\it shift} and {\it shiftLocal} behave similarly if not better than the rest of the techniques.

The normalized results are summarized in Tables \ref{Table:exp1-time} and \ref{Table:exp1-time64}, where the value 1 corresponds to the throughput of the no balance technique in each setting.  When the imbalance is high, {\it probCheck} dominates all the other rebalancing options in terms of performance for small grid sizes. In those cases, the speed up is more than 4 times. When the grid size is increased, the best performing technique is {\it getFirst}, which nearly doubles the throughput. So, we can deduce that less sophisticated and approximate load balancing techniques, which require less computational effort for the balancing itself, are more appropriate for GPGPU in highly skewed environments. When the imbalance is low, the maximum speedup is negligible for grid size set to 4, and {\it shift} and {\it shiftLocal} seem more appropriate. When we further increase the grid size, the throughput may degrade. We also discuss the impact of the grid size in more detail later.

Finally, as mentioned before, the time spent in load balancing on the CPU may be hidden by the time spent in GPU execution and data transfer. This holds for the case where the grid size is set to 4. When we set the grid size to 64, the CPU-based data preparation partially overlaps with the GPU-based aggregate computation, and, as a result, the preparation overhead affects the performance.

\subsubsection{Overhead}

To further investigate the overhead incurred by maintaining the auxiliary structures and performing the rebalancing computations, in Figure \ref{fig:DS1}, we show the increase in the total time required when we let the skew handling techniques to unnecessarily investigate possible rebalancing  actions; to this end we use DS1. From the figure, we can observe that simply enabling any of the methods has an impact on the performance, which is more evident with increased grid sizes. Actually, the overhead is negligible for the small grid size, where the CPU cost is fully hidden by the GPU processing. In addition, {\it shiftLocal} is the technique with almost negligible overhead for any combination of system and grid size; this is attributed to the fact that it does not utilize heap structures for finding the most and least loaded groups.

\subsubsection{More on the effect of grid size}
The grid size, as evidenced in all the performance figures thus far, is an important parameter that affects the performance.
By increasing the grid size, we increase the number of threads produced in the card. Consequently, the number of groups to be processed by a thread is reduced. As a result, the odds of having a thread, which gathers many overloaded groups decrease; in other words, imbalance effects are indirectly mitigated through increased grid sizes. Thus, a larger grid size leads to shorter processing time, as much fewer threads will run for very long time acting as a bottleneck. Taking advantage of the very low thread scheduling time offered by the Fermi architecture, we experience reduced total times, even without any of the skew handling techniques enabled. However, increasing the grid size makes sense only when there is at least one group in the aggregate query per thread. In our case, where we have 40K groups, this has proved to be not a problem, but in several other scenarios the total number of groups in the aggregate query may be a few hundreds, which necessitates smaller grid sizes because larger grid sizes are simply not applicable in those scenarios.

\begin{figure}[tb!]
\begin{center}
\includegraphics[width=0.49\textwidth]{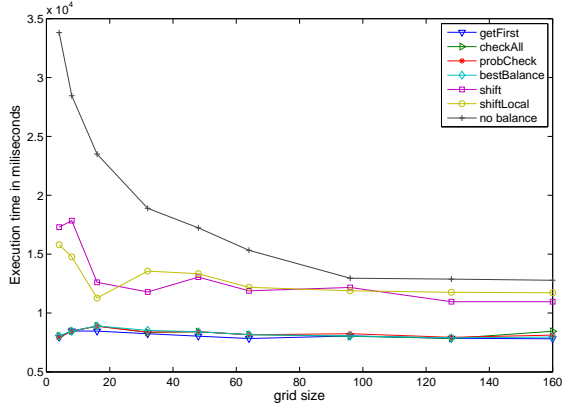}
\caption{The effect of changing the gridsize in DS2 (PC1).}
\label{fig:gridsize}
\end{center}
\end{figure}

Moreover, if we increase the number of threads over a certain point, we reach a saturation point where there is no more benefit. This is due to the fact that the CPU, which maintains the structures that keep data for the thread load factor, now needs more time to process them. Consequently, the total execution time cannot be reduced under a limit, and in some cases the additional CPU effort might outweigh the benefit. Only in the case of {\it shiftLocal}, which does not need those extra structures and computations, the surcharge is small. Additionally, we notice that the {\it shift} method also benefits from the decrease of groups per thread due to increased grid size, as it has higher chances to achieve acceptable balance.

Figure \ref{fig:gridsize} shows the impact of the grid size on the running time for DS2 being executed on PC1. We can observe that, by increasing the grid size, we can alleviate high imbalances, although we cannot eliminate them, since the difference between no balancing and balancing schemes decreases significantly but it never becomes negligible. For {\it shift} and {\it shiftLocal}, the saturation point is much earlier, whereas all the other techniques are insensitive to the grid size.

\subsubsection{CPU vs GPU}
\begin{figure}[tb!]
\begin{center}
\includegraphics[width=0.45\textwidth]{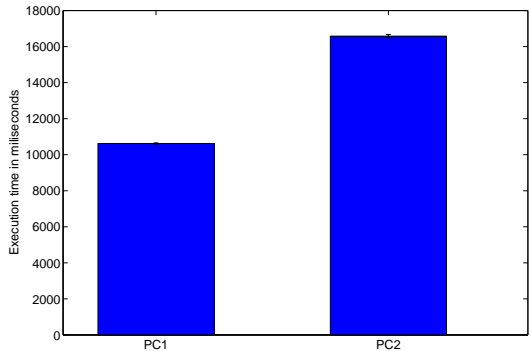}
\caption{Execution time in CPU.}
\label{fig:cputimes}
\end{center}
\end{figure}

Additionally, in order to demonstrate the power of the GPU processing, we implemented a simple CPU-based group by algorithm. There, as the data is not distributed among threads but handled serially, no skew handling is needed and the total throughput is the same in all datasets of equal size. The execution times are presented in the Figure \ref{fig:cputimes}. Compared with the running times in Figures \ref{fig:DS2}-\ref{fig:DS1}, it is evident that employing the GPU is always beneficial. For example, in DS2, the running time using the GPU can be as low as approximately 8 secs for both systems, whereas it is more than 10 and 16 secs when we use only the CPU of PC1 and PC2, respectively. The performance benefits are larger for DS1 and DS3.

\subsubsection{Further experiments and discussion}

\begin{figure}[tb!]
\begin{center}
\includegraphics[width=0.49\textwidth]{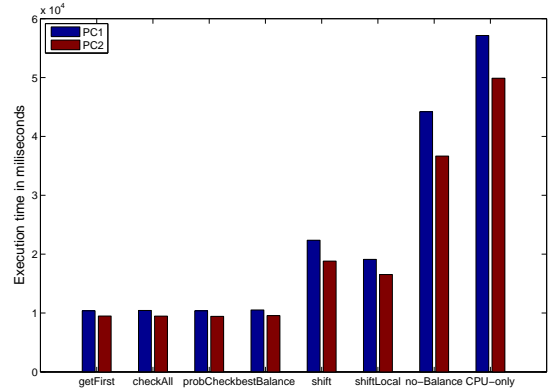}
\caption{Performance comparison when there is a 10-fold increase in the window passes for DS2 (the grid size is 4).}
\label{fig:10passes}
\end{center}
\end{figure}

In our last experiment, we investigate adding some extra load: we assume an aggregate function that checks the window elements 10 times instead of one per window update.  Despite the fact that modern graphics cards are alleged to perform much better than the CPU on processing numeric and double precision data, rather than on accessing data from the card's global memory for a simple summing computation, their superiority and capability to yield improved performance is obvious in Figure \ref{fig:10passes}.

Finally, as evidenced in the figures, there is no clear winner between the two systems. PC2, owing to the faster and more powerful graphics processing unit, proved to be faster in some cases were the majority of computation happens in the GPU. DS2 is such a case, where the imbalance among the threads acts as a bottleneck, and the CPU processing time is negligible compared to the GPU processing. Also, PC2 benefits from the smaller grid sizes that require less CPU effort in order to handle the corresponding structures. Meanwhile, PC1 outperforms PC2 in the case of bigger grid sizes, where the need for skew handling is decreased, as explained earlier.

\section{Conclusions}
\label{sec:conclusions}

In this work, we dealt with the problem of skewed execution in aggregate queries on GPUs. We presented a generic load balancing framework along with specific balancing techniques. The lessons learnt can be summarized as follows: Firstly, we can significantly reduce execution time with the help of GPUs. In our experiments, we observed significant speed-ups up to 4 times and verified the fact that load imbalances can lead to serious performance degradations thus necessitating load balancing actions in order to avoid performance degradation. Secondly, the techniques we proposed are both efficient and effective; their efficiency is due to their low overhead, whereas their effectiveness is manifested through their capability to lead to throughput improvements in highly skewed scenarios. Interestingly, less sophisticated and approximate techniques exhibit superior performance, because the increased overhead of more sophisticated solutions outweighs any benefits and/or load balancing decisions are enforced with a delay of one round, where the exact conditions may have changed anyway. When small imbalances are experienced, no balancing technique behaves well. Finally, increasing the grid size mitigates the effect of skewed executions, but it is not always applicable in aggregate queries. Overall, this work aims to provide useful insights into the behaviour of rebalancing techniques for GPU-assisted data management.

Potential avenues for future work include the investigation of integrated techniques that both balance the load among the threads and vary the grid and the batch size, and of orthogonal issues, such as more efficient methods of data exchange between the GPU and the main memory. Finally, it is worth investigating the behaviour of our approach in the light of the recent dynamic parallelism extensions of the CUDA programming model, introduced with the Kepler architecture.

\section{Acknowledgments}
This research has been co-financed by
the European Union (European Social Fund - ESF) and
Greek national funds through the Operational Program ``Education and Lifelong Learning" of the National Strategic
Reference Framework (NSRF) - Research Funding Program:
Thales. Investing in knowledge society through the European Social Fund.

\balance

\bibliographystyle{abbrv}
\bibliography{gpubib}

\end{document}